\magnification=1200
\baselineskip=22pt plus 1pt minus 1pt
\tolerance=1000
\parskip=0pt
\parindent=25pt



\font\bxBig=cmbx10  scaled\magstep3

\font\medsl=cmsl8

\def\normal{\baselineskip=22pt plus 1pt minus 1pt}

\def\ep{\epsilon}
\def\eps{\varepsilon^{\mu\nu\rho}}
\def\d{\partial}
\def\la{\raise.16ex\hbox{$\langle$} }
\def\ra{\raise.16ex\hbox{$\rangle$} }
\def\go{\rightarrow}

\def\psibar{ \psi \kern-.65em\raise.6em\hbox{$-$} }
\def\Dbar{ D \kern-.8em\raise.65em\hbox{$-$} }

\def\bc{ {\bar c}}
\def\bJ{ {\bar J}}
\def\bq{{ \bar q}}
\def\bz{ {\bar z}}

\def\N{ \kappa }

\def\m{ {\bf m} }
\def\n{ {\bf n} }

\nopagenumbers
\rightline{Aug, 1995}
\rightline{Revised Version}
\vskip 1.8cm

\baselineskip=22pt
\centerline{\bxBig W$_{\infty}$ and SL$_q$(2) Algebras in the Landau }
\centerline{\bxBig  Problem and Chern-Simons Theory on a Torus}

\vskip 1.8cm

\baselineskip=13pt

\centerline{\bf Choon-Lin Ho}
\centerline{\medsl Department of Physics, Tamkang University, Tamsui,
Taiwan 25137, R.O.C.}

\vskip 1.5cm

\baselineskip=16pt

\centerline{\bf Abstract}

\midinsert \narrower

We discuss $w_\infty$ and $sl_q(2)$ symmetries in Chern-Simons theory and
Landau problem on a torus.  It is shown that when the coefficient of the
Chern-Simons term, or when the total flux passing through the torus is
a rational number,
there exist in general two $w_\infty$ and two $sl_q(2)$ algebras,
instead of
one set each discussed in the literature.  The general wavefunctions for
the Landau problem with rational total flux is also presented.

\endinsert

\vfill\eject

\footline={\hss\tenrm\folio\hss}


\pageno=2

\normal
\parindent=25pt

Chern-Simons (CS) field theory with matter coupling have attracted
intense
interest in recent years, owing to its relevance to condensed matter
systems
such as  quantum Hall systems, and possibly  high $T_c$
superconductivity.
While the majority of works in the field is concerned with planar
systems,
Chern-Simons field theory on compact Riemann surfaces has also captured
considerable interests.  It has even richer structures, which, being
topological in nature, are absent in planar system.  Among them are the
multicomponent structure of  many-body wavefunctions and the
degeneracy of physical states (see references in [1]).

Many studies have also been carried out for the Maxwell-Chern-Simons
(MCS) theory
in which a kinetic term for the CS gauge field is included.  An
interesting
observation is that the dynamics in the topological sector of MCS
theory on a
torus is equivalent to the Landau problem ({\it i.e.} charged
particle moving
in a constant magnetic field) on a torus [2-4].  Thus many interesting
features are shared by both systems.

Recently a  $w_\infty$ symmetry is uncovered
in the Landau problem and in the related problem of fractional quantum Hall
effects [5-7].  More recently, a $sl_q(2)$ quantum
algebra is realized  in these systems [8,6{\it b},9].
Representation of this quantum algebra was applied to formulate the
Bethe-{\sl ansatz}
for the problem of Bloch electron in magnetic field, {\it i.e.} the
Azbel-Hofstadter problem [8].  Naturally, these symmetries were realized
also in the MCS and the pure CS theory on a torus [6{\it b}].

The $w_\infty$ and $sl_q(2)$ algebras considered in these works are mainly
for the situations
in which the total flux passing through the torus (or unit cell in the
lattice)
is an integral multiple of $2\pi$, or equivalently in the CS theory
when the
CS coefficient is an integer.  In this paper, we would like to point
out that
in the more general case, that is, when the flux or the CS coefficient is
rational, there could exist two $w_\infty$ and two $sl_q(2)$ algebras.

Let us consider a Maxwell-Chern-Simons (MCS) theory defined on a torus
with lengths $L_1$ and $L_2$.
The Lagrangian is given by
$$ L= -{\alpha\over 4}f_{\mu\nu}^2 + {\N \over 4\pi} ~\eps a_\mu \d_\nu
a_\rho~. \eqno(1) $$
Here we assume $\N$ to be a rational number, $\N=N/M$, where $N$, $M$ are
two coprime integers.   The equations of motion can be solved in the
$a_0=0$ gauge
[3,4]. The spatial components $a_j$ are found to decompose into global
excitations, $\theta_i(t)/L_i$, and local excitations.  Contributions of
these two parts decouple in the action of the theory.  The global
excitations
$\theta_i$ are the non-integrable phases associated with the two
non-contractible loops of the torus.  They are responsible for the
topological
structures of the theory.  In the rest of the paper, we will consider the
structures of the Hilbert space of the global excitations.

The Lagrangian of the $\theta_i$'s is given by
$${\cal L} = {\alpha\over 2}\left({L_2\over L_1}{\dot \theta_1}^2 +
{L_1\over L_2}{\dot \theta_2}^2\right) +
{\N\over 4\pi}\left(\theta_2{\dot\theta_1} -
\theta_1{\dot\theta_2}\right),\eqno(2)$$
from which the corresponding canonical momenta are obtained:
$$\eqalign{
p_1 &= \alpha {L_2\over L_1} {\dot \theta_1} + {\N\over 4\pi}\theta_2~,\cr
p_2 &= \alpha {L_1\over L_2} {\dot \theta_2} - {\N\over 4\pi}\theta_1~,\cr
&[\theta_i, p_k]=i\delta_{ik}~.\cr}
\eqno(3)$$

As mentioned previously, the system defined by (2) is equivalent to the
Landau problem on a torus.
In fact, if one defines $x_i\equiv \theta_i/L_i$, $L_{x_i}\equiv
2\pi/L_i$, $m\equiv \alpha L_1L_2$ and $B\equiv 2\pi\N/L_{x_1}L_{x_2}$,
then the Hamiltonian governing the dynamics is
$$H={1\over 2m} \left[\left(p_{x_1}-{1\over 2}Bx_2\right)^2
     + \left(p_{x_2}+{1\over 2}Bx_1\right)^2 \right]~,\eqno(4)$$
where $p_{x_i} = m{\dot x}_i + {1\over 2}\ep^{ik}B x_k$,
$[x_i,p_{x_k}]=i\delta_{ik}$.  This is just the
Hamiltonian for a charged particle with mass $m$ moving in constant
magnetic
field $B$ perpendicular to the torus in the symmetric gauge. The total
flux on the torus is
$\Phi=BL_{x_1}L_{x_2}=2\pi\N$.  The energy spectrum and wavefunctions
of this problem has been considered before for the case when $\N=N$
($M=1$) [3,10-13].
It is found that the Landau levels
are $N$-fold degenerate.  Wen and Niu [3] consider the case $\N=1/M$ in
connection
with the problem of Fractional Quantum Effect, and show that the
degeneracy of states is
$M$.  One of the purpose of this paper is to show that in the general case
$\N=N/M$, the Landau levels are $MN$-fold degenerate, and to present an
explicit expression for these wavefunctions.

To construct the wavefunctions, one has to first identify the appropriate
symmetry algebra of the system.  The wavefunctions must form a
representation
of the algebra.  For the Landau problem, the relevant algebra is the
Weyl-Heisenberg (WH) algebra of the so-called magnetic translation
operators
(MTO) [14].  What we would like to emphasize here is that in general
there are two
commuting  WH algebras in the system, and not just one as usually
considered in the
literature.  One of these two algebras leads to an $N$-fold
degeneracy, while
the other to a $M$-fold degeneracy, giving rise to a total of $MN$
degenerate states.

The MTO in this problem are given by  [14] (no summation over repeated
indices)
$$U_j=\exp\left\{L_{x_j}\left(ip_{x_j}
        + {i\over 2}\ep^{jk}Bx_k\right)\right\}~.
\eqno(5)$$
In the original MCS theory they are the generators of
large gauge transformations inducing
$\theta_j \go \theta_j + 2\pi$:
$$U_j=\exp\left(2\pi ip_j + {i\over 2}\N \ep^{jk}\theta_k\right)~.
\eqno(6)$$
They commute with the Hamiltonian (4).  Now there is another set of
operators
which commute with $H$ and $U_j$.  These operators are defined by
$V_j \equiv U_j^{1/\N}$.
$U_j$ and $V_j$ satisfy dual WH algebras :
$$\eqalign{
U_1U_2 &= e^{-2\pi i \N}\, U_2U_1~,\cr
V_1V_2 &= e^{-2\pi i/ \N} \,V_2V_1~, \cr}\eqno(7)$$
and $[U_j, V_k]= 0$.
The $V_j$ have sometimes been loosely called ``magnetic translation
operators"
in some papers (And this, we believe, does cause some confusion).
Now the eigenfunctions $\psi$ of $H$ must form a representation of (7).
In previous works on Landau problem on a torus, only one of the two WH
algebras
in (7) were employed in the study of the structure of Hilbert space.
To get a
complete Hilbert space when the total flux through the torus is rational,
one must consider both algebras.
By noting that $U_1^M, U_2^M, V_1^N$ and $V_2^N$ commute with each
other, we can choose $\psi$ to be eigenstates of $U_i^M$ and $V_j^N$ :
$U_i^M \psi = V_i^N \psi =e^{i\alpha_i} \psi $ ($i=1,2$), where
$\alpha_i$ are
the vacuum angles.
The problem is more easily solved in terms of complex variables.
Following the procedure in [11,12],  we define
$u_a=\sqrt{B} x_a$, $l_a=\sqrt{B} L_{x_a}~(l_1l_1=2\pi\N)$ and
$z=u_1+iu_2$.
Then the Hamiltonian is
$H={B\over m} \left(\bc c +{1\over 2}\right)$, where
$c\equiv i\sqrt{2}\left(\partial_z + {1\over 4}\bz\right)$,
$\bc\equiv i\sqrt{2}\left(\partial_\bz - {1\over 4} z\right)$ and
$[c,\bc]=1$.  Hence the energy spectrum is $E={B\over m}\left(n+{1\over
2}\right), n=0,1,2,\ldots$.  The wavefunctions of Landau levels are
given by $\psi_n (z,\bz) = \bc^n \psi_0 (z,\bz)$, where the ground state
$\psi_0$ satisfies
$c\psi_0=0$.  The problem of finding the general wavefunctions now reduces to
that of finding $\psi_0$.  To this end, we note that the requirements
$U_i^M \psi =e^{i\alpha_i} \psi $ stated above imply the following
boundary conditions of $\psi$ in complex variable form:
$$\eqalign{
\psi (z+Ml_1, \bz+Ml_1) &=e^{-{1\over 4}Ml_1 (z-\bz)+i\alpha_1}~\psi
(z,\bz)~,\cr
\psi (z+iMl_2, \bz-iMl_2) &=e^{+{1\over 4} i Ml_2 (z+\bz)+i\alpha_2}~\psi
(z,\bz)~.\cr}\eqno(8)$$
We now let $\psi_0$ assume the form:
$$\psi_0 (z,\bz)=\xi (z,\bz)~G(\bz)~,\eqno(9)$$
where $G(\bz)$ is a function of $\bz$ only.  Then the condition $c\psi_0=0$
implies $c\xi=0$, which is solved by
$$\xi = \exp\left(-{1\over 4}z\bz + {1\over 4}\bz^2
+ i\delta_1\bz\right)~~,~\delta_j\equiv \alpha_j/Ml_j~.\eqno(10)$$
From (9) and (10) one finds that the function $G$ obeys the
following boundary conditions
$$\eqalign{
G(\bz+Ml_1) &= G(\bz)~,\cr
G(\bz-iMl_2) &= e^{{1\over 2} (Ml_2)^2 + i Ml_2(\bz+i\delta)} ~G(\bz)~,\cr}
\eqno(11)$$
where $\delta = \delta_1 - i\delta_2 $.  Hence $G$ is just a doubly periodic
function  related to the theta functions. Since the area bounded by
the two translations $Ml_1$ and $Ml_2$ on the $z$ plane is $M^2l_1l_2 =
2\pi MN$ (recall $l_1l_2=2\pi\kappa$),  there exist $MN$ independent
solutions of $G$.  These functions are conveniently labeled by two
indices $j,k$ ($j=0,\ldots,N-1;~k=0,\ldots,M-1$)
and are given by
$$
G_{jk}(\bz)=\sum_{n=-\infty}^{\infty} Q^{(MNn-Nk-Mj)^2\over MN}
e^{-2\pi i (MNn-Nk-Mj)(\bz + i\delta)/Ml_1}~. \eqno(12)$$
Here $Q\equiv e^{-\pi l_2/l_1}$ .  So each Landau level is $MN$-fold
degenerate : $\psi_{njk}= \bc^n \psi_{ojk}$ and $\psi_{0jk}(z,\bz) =
\xi(z,\bz)~G_{jk}(\bz)$.
The function $G_{jk}(\bz)$ is
expressible in terms of the theta functions with characteristics:
$$\eqalign{
G_{jk}(z,\bz) &= \vartheta \left[\matrix{-{k\over M}-{j\over N}\cr
                                -i{N\over l_1}\delta\cr}\right]
 ~\left(-{N\over l_1}\bz~,~iMN{l_2\over l_1}\right)~,\cr
\vartheta \left[\matrix{a\cr b\cr}\right]\left(z,\tau\right) &=
\sum_n \exp\left(i\pi\left(n+a\right)^2 \tau + 2\pi
i\left(n+a\right)\left(z+b\right)\right)~.\cr}\eqno(13)$$.

We now consider the actions of the $U_j, ~V_j$ on $\psi_{njk}$.
To simplify discussions, we will concentrate only on the first
Landau level $n=0$.  The discussion given below can be carried over
directly to higher levels, since the $U_j$ and $V_j$ commute
with $c$ and $\bc$. So let us denote by $\langle z,\bz|jk\rangle\equiv
\psi_{0jk}(z,\bz)$.  The actions of $U_i, ~V_i$ on the state vectors are :
$$\eqalign{
U_1|jk\rangle &= e^{i(\alpha_1 + 2\pi N k)/M} |jk\rangle~~,\cr
U_2|jk\rangle &= e^{i\alpha_2/M} |j,k-1\rangle~~,\cr
V_1 |jk\rangle &=e^{i(\alpha_1 + 2\pi M j)/N} |jk\rangle~~,\cr
V_2 |jk\rangle &=e^{i\alpha_2/N} |j-1,k\rangle~~.\cr}
            \eqno(14)$$

In the limit $\alpha\rightarrow 0$, the MCS theory reduces to the
pure CS theory.  This corresponds to the reduction of the
Hilbert space of the Landau problem to the first level.  The Lagrangian
now becomes ${\cal L}={\N\over 4\pi}\left(\theta_2{\dot\theta_1} -
\theta_1{\dot\theta_2}\right)$ and $H$ vanishes identically.  From the
Lagrangian
one sees that $\theta_1$ and $\theta_2$ become canonical variables of the
reduced phase space , and thus $[\theta_1,~\theta_2]=2\pi
i/\N$.  The Hilbert space structure of this theory is well studied
[1-4,15-17]. The operators $U_j$ now reduce to the form
$U_j =\exp \big\{ i\ep^{jk} \N \, \theta_k \big\}$, which is still the
generators of large gauge transfomations in the theory.  For the $V_j$,
one has
$(V_1,V_2)\rightarrow (e^{i\theta_2}, e^{-i\theta_1})\equiv (W_2,
W_1^{-1})$.
The operators $W_j$ are nothing but the Wilson line operators.  It
is easy to check that $W_j$ satisfy the same WH algebra as the $V_j$:
$W_1W_2 = e^{-2\pi i/ \N} \,W_2W_1$ and  $[W_k, U_j]= 0$.  As before, the
wavefunctions must form a
representation of the two WH algebras.   To comform to the expressions
given in previous work [1],
we shall work in the
$\theta_1$-representation and diagonalize
both $U_1$ and $W_1$ .  The resulted states $u_{jk}$ are $NM$-degenerate :
$u_{jk}(\theta_1)= \la \theta_1 |
jk\ra  =e^{ik(\alpha_2 + N\theta_1)/M + i \alpha_1 \theta_1/2\pi M } ~
 \delta_{2\pi} [\theta_1 + (\alpha_2- 2\pi Mj)/N]$.
Actions of the $U_i, W_j$ on $|jk\rangle$ are easily found to be given by
(14), except that in the third and fourth expresions, the $V_j$ are
replaced by $W_j$, and $(\alpha_1,\alpha_2)$ are
replaced by  $(-\alpha_2,\alpha_1)$
(recall that $V_1\rightarrow W_2$ and
$V_2\rightarrow W_1^{-1}$) [1].

We now turn to discuss the quantum $w_\infty$-symmetries and quantum
algebras
hidden in the MCS and the Landau problem.  It is now well known that the
magnetic translation operators, or to be more precise, the generators
of the
WH algebra, span the quantum $w_\infty$ algebra (also termed FFZ algebra
[18]) [5,6,9{\it a}], and can induce a quantum symmetry $sl_q(2)$, where
$q$ is the deformed parameter
[8,6,9].  However, only one set of WH algebra was
considered in the
works just cited.  Particularly, in [6] and [9{\it b}] only the case for
which the
total flux through the torus is $\Phi=2\pi N$ ($M=1$) is considered.  As
we see
before, there are in general two WH algebras present when $\Phi=2\pi N/M$
and
$N\neq M \neq 1$.  One thus expect to find two quantum $w_\infty$ algebras
 and
two $sl_q(2)$.  This is indeed the case.  Below we construct the generators
for
these algebras in the MCS theory (which is the same as the Landau problem).
The entire results can be easily
extended to the pure CS theory by making the appropriate changes in $V_j$,
$W_j$ and $\alpha_j$ mentioned previously.

Let us define the following operators [5,19] :
$$\eqalign{
S_\n &=S_{(n_1,n_2)}\equiv q^{n_1n_2/2} U_1^{n_1} U_2^{n_2} ~,\cr
T_\n &=T_{(n_1,n_2)}\equiv \bq^{n_1n_2/2} V_1^{n_1} V_2^{n_2} ~,\cr}
\eqno(15)$$
where $n_1, n_2$ are integers , $q\equiv e^{2\pi i\N} ~(q^M=1)$ and
$\bq\equiv e^{2\pi i/\N} ~(\bq^N=1)$.  $S_\n$ and $T_\n$ are simply
operators that generate general magnetic translations on the torus.
Using the WH algebras (7) one obtains :
$$\eqalign{
S_\m~S_\n = q^{-\m\times \n/2} S_{\m+\n}~,\cr
T_\m~T_\n = \bq^{-\m\times \n/2} T_{\m+\n}~.\cr}\eqno(16)$$
Here $\m\times \n=m_1n_2-m_2n_1$.
From (16) one easily gets the two quntum $w_\infty$ mentioned before:
$$\eqalign{
\left[S_\m,S_\n\right] & = -2i
\sin\Bigl(i\pi\N\left(\m\times\n\right)\Bigr)S_{\m+\n}~,\cr
\left[T_\m,T_\n\right] & = -2i
\sin\left({i\pi\over \N}\left(\m\times\n\right)\right)T_{\m+\n}~,\cr}
\eqno(17)$$
Only the $T$-algbra was given in [6].  Using (14), the
actions of $S_\n$ and $T_\n$ on the state $|jk\rangle $ are found to be:
$$\eqalign{
S_\n |jk\rangle &=q^{n_1n_2\over 2} e^{i\alpha_2{n_2\over M}}
               e^{in_1[\alpha_1 + 2\pi (k-n_2)N]/M} |j,k-n_2 \rangle~\cr
T_\n |jk\rangle &=\bq^{n_1n_2\over 2} e^{i\alpha_2{n_2\over N}}
               e^{in_1[\alpha_1 + 2\pi (j-n_2)M]/N} |j-n_2,k \rangle~.\cr}
\eqno(18)$$
When $\alpha_i=0$ and $M=1$, the expression for $T_\n$ reduces to eq.(3.19)
in [6{\it b}] (note that $\N/2$ in [6{\it b}] is equivalent to $\N$ here).

From the  operators (15) one can construct two quantum algebras
$sl_q(2)$ and $sl_\bq(2)$.  The generators of a general $sl_q(2)$ are
defined by:
$$\eqalign{
q^{J_3}J_\pm q^{-J_3} &=q^{\pm 1}J_\pm~,\cr
[J_+,J_-]&=[2J_3]_q~,\cr}\eqno(19)$$
where $[x]\equiv (q^x-q^{-x})/(q-q^{-1})$.
In our case, the $J$'s for the $sl_q(2)$ are constructed from the $S_\n$
as follows [6,9]:
$$\eqalign{
J_+ &\equiv {1\over q-q^{-1}}\left(S_{(1,1)} - S_{(-1,1)}\right)~,\cr
J_- &\equiv {1\over q-q^{-1}}\left(S_{(-1,-1)} - S_{(1,-1)}\right)~,\cr
q^{2J_3} &\equiv S_{(-2,0)}~, \qquad\qquad
q^{-2J_3} \equiv  S_{(2,0)}~.\cr}\eqno(20)$$
The generators $\bJ$'s for the $sl_\bq(2)$ algebra are also constructed
in the
same way as in (20), with $J, q, S$ replaced by $\bJ,\bq, T$ respectively.

With the help of (18), it is easy to obtain the actions of the $J,\bJ$
on the state vectors $|jk\rangle$ :
$$\eqalign{
J_+ |jk\rangle &= e^{i\alpha_2/M} \left[k-{1\over 2}+{\alpha_1\over 2\pi
N}\right]_q |j, k-1\rangle~,\cr
J_- |jk\rangle &=- e^{-i\alpha_2/M} \left[k+{1\over 2}+{\alpha_1\over 2\pi
N}\right]_q |j, k+1\rangle~,\cr
q^{\pm 2J_3} |jk\rangle &=q^{\mp 2\left(k+{\alpha_1\over 2\pi N}\right)}
|jk\rangle~,\cr}\eqno(21)$$
and
$$\eqalign{
\bJ_+ |jk\rangle &= e^{i\alpha_2/N} \left[j-{1\over 2}+{\alpha_1\over 2\pi
M}\right]_\bq |j-1, k\rangle~,\cr
\bJ_- |jk\rangle &=- e^{-i\alpha_2/N} \left[j+{1\over 2}+{\alpha_1\over 2\pi
M}\right]_\bq |j+1, k\rangle~,\cr
\bq^{\pm 2\bJ_3} |jk\rangle &=\bq^{\mp 2\left(j+{\alpha_1\over 2\pi M}
\right)} |jk\rangle~,\cr}\eqno(22)$$

In view of (14) the state functions $|jk\rangle$ in general form a cyclic
representation of $sl_q(2)\times sl_\bq (2)$ with dimension $M\times N$.
 However, under certain choices  of the boundary conditions $\alpha_i$,
highest weight representation can be formed [20].  For simplicity, let us
first give a general consideration for the subspace $|k\rangle
= |j,k\rangle$ acted by the  $sl_q(2)$  algebra ($q^M=1$).  Now
$|k+M\rangle=|k\rangle$ implies
that the states form a $M$-dimensional cyclic representation of $sl_q(2)$
in general.  But from
(21), and the observation that $[{M\over 2}]_q=0$, one concludes that
if the
boundary condition $\lambda_N\equiv \alpha_1/2\pi N$ is such that
$l={M\over 2} - \left(\lambda_N+{1\over 2}\right)$ is an integer, then the
state
$|l+1\rangle$ is the highest and $|l\rangle$ is the lowest weight states:
$$\eqalign{
&J_+ |l+1\rangle \sim \left[{M\over 2}\right]_q |l\rangle =0~,\cr
&J_- |l\rangle \sim -\left[{M\over 2}\right]_q |l+1\rangle =0~.\cr}
\eqno(23)$$
The succesive actions of $J_-$ starting from the highest weight state is:
$|l+1\rangle \rightarrow \cdots \rightarrow |M\rangle=|0\rangle \rightarrow
|1\rangle \rightarrow\cdots \rightarrow |l\rangle~$
($J_+$ acts reversely).
The condition that $l$ be an integer requires that $\lambda_N=m$
for odd $M$, and $\lambda_N+{1\over 2}=m$ for even $M$ ($m=$integers).
Similar consideration leads to the same requirements for $\lambda_M\equiv
\alpha_1/2\pi M$ for the $sl_\bq(2)$ algebra :
$\lambda_M  (\lambda_M+{1\over 2})=
{\rm integers } $ for odd (even) $N$.  Combining these considerations for
both sectors, one concludes that the states $|j,k\rangle$ can form
highest weight representation for both $sl_q(2)$ and $sl_\bq(2)$ algebras
only when $M$ and $N$ are both odd, and
$\alpha_1=n\times MN$ where $n$ are integers.  There is no restriction on
the
$\alpha_2$.  This unbalanced requirement in the two vacuum angles is only
due to our diagonalizing both $U_1$ and $V_1$.  The situation changes
if we diagonalize
instead $U_1$ and $W_1=V_2^{-1}$ as in the pure CS theory, then both vacuum
angles
are constrained in order to have highest weight representation for both
algebras.
In this case, $\alpha_1/2\pi N ~(\alpha_2/2\pi M) =m~{\rm or}\
m-{1\over 2}$ if $M~(N)$ is odd or even respectively.

\vskip 1. truecm

\centerline{\bf Acknowledgement}

This work is supported by R.O.C. Grant NSC 84-2112-M032-001.

\vfil\eject

\def\ap#1#2#3{{\it Ann.\ Phys.\ (N.Y.)} {\bf {#1}}, #3 (19{#2})}
\def\cmp#1#2#3{{\it Comm.\ Math.\ Phys.} {\bf {#1}}, #3 (19{#2})}
\def\ijmpA#1#2#3{{\it Int.\ J.\ Mod.\ Phys.} {\bf {A#1}}, #3 (19{#2})}
\def\ijmpB#1#2#3{{\it Int.\ J.\ Mod.\ Phys.} {\bf {B#1}}, #3 (19{#2})}
\def\jmp#1#2#3{{\it  J.\ Math.\ Phys.} {\bf {#1}}, #3 (19{#2})}
\def\mplA#1#2#3{{\it Mod.\ Phys.\ Lett.} {\bf A{#1}}, #3 (19{#2})}

\def\plB#1#2#3{{\it Phys.\ Lett.} {\bf {#1}B}, #3 (19{#2})}

\def\np#1#2#3{{\it Nucl.\ Phys.} {\bf B{#1}}, #3 (19{#2})}
\def\prl#1#2#3{{\it Phys.\ Rev.\ Lett.} {\bf #1}, #3 (19{#2})}
\def\pr#1#2#3{{\it Phys.\ Rev.} {\bf {#1}}, #3 (19{#2})}
\def\prB#1#2#3{{\it Phys.\ Rev.} {\bf B{#1}}, #3 (19{#2})}

\def\jpA#1#2#3{{\it J.\ Phys.} {\bf A{#1}}, #3 (19{#2})}
\def\ap#1#2#3{{\it Ann.\ Phys.\ (N.Y.)} {\bf {#1}B}, #3 (19{#2})}
\def\jetp#1#2#3{{\it Sov.\ Phys.\ JETP} {\bf {#1}}, #3 (19{#2})}
\def\cnpp#1#2#3{{\it Comm.\ Nucl.\ Part.\ Phys.} {\bf {#1}}, #3 (19{#2})}
\def\ptp#1#2#3{{\it Prog.\ Theor.\ Phys.} {\bf {#1}}, #3 (19{#2})}

\parindent=15pt

\centerline{\bf References}

\item{[1]}
C.-L. Ho and Y. Hosotani, \ijmpA {7} {92} {5797}; \prl {70} {93} {1360};
D. Wesolowski, Y. Hosotani and C.-L. Ho, \ijmpA {9} {94} {969}.

\item{[2]}
I.I.Kogan and A. Yu. Morozov, \jetp {61} {85} {1};
I.I. Kogan, ITEP preprint ITEP-89-163 (1989); \cnpp {19} {90} {305}.

\item{[3]}
X.G. Wen, \prB {40} {89} {7387}; \ijmpB {4} {90} {239};
X.G. Wen and Q. Niu, \prB {41} {90} {9377}.

\item{[4]}
A.P. Polychronakos, \ap{203}{90}{231}; Univ. of Florida preprint,
 UFIFT-HEP-89-9;
 \plB {241} {90} {37}.

\item{[5]}
T. Dereli and A. Vercin, \plB {288} {92} {109}; \jpA {26} {93} {6961}.

\item{[6]}
I.I. Kogan, a)\mplA {7} {92} {3717}; b) \ijmpA {9} {94} {3889}.

\item{[7]}
S. Iso, D. Karabali and B. Sakita, \plB {296} {92} {143};
B. Sakita, \plB {315} {93} {124};
A. Cappelli, C.A. Trugenberger and G.R. Zembra, \np {396} {93} {465};
A. Cappelli, G.V. Dunne, C.A. Trugenberger and G.R. Zembra, \np {398} {93}
{531}.

\item{[8]}
P.B. Wiegmann and A.V. Zabrodin, cond-mat/9310017; cond-mat/9312088;
L.D. Faddeev and R.M. Kashaev, hep-th/9312133.

\item{[9]}
H.-T. Sato,a) \mplA {9} {94} {451}; b) {\it ibid.} 1819;
\ptp  {93} {94} {195}.

\item{[10]}
R.B. Laughlin, \ap {191} {89} {163}.

\item{[11]}
C.-L. Ho, \mplA {5} {90} {2029}.

\item{[12]}
M. Burgess, A. McLachlan and D.J. Toms, \ijmpA {8} {93} {2623}.

\item{[13]}
For Landau problem in higher genus Riemann surfaces, see:
R. Iengo and D. Li, \np {413} {94} {735}.

\item{[14]}
E. Brown, \pr {133} {63} {1038};
J. Zak, \pr {134} {64} {1602}; {\it ibid.} 1607.

\item{[15]}
Y. Hosotani, \prl {62} {89} {2785};  \prl {64} {90} {1691}.

\item{[16]}
K. Lee, Boston Univ. report,   BU/HEP-89-28;

\item{[17]}
R. Iengo and K. Lechner, \np {346} {90} {551}; \np {364} {91} {551};
K. Lechner, \plB {273} {91} {463}; Trieste SISSA preprint, Thesis,
Apr 1991.

\item{[18]}
D.B. Fairlie, P. Fletcher and C.K. Zachos, \plB {218} {89} {203};
\jmp {31} {90} {1088}.

\item{[19]}
M. Golenishcheva-Kutuzova and D. Lebedev, \cmp {148} {92} {403};
J. Hoppe, M. Olshanetsky and S. Theisen, \cmp {155} {93} {429}.

\item{[20]}
For an extensive discussion on the r\^oles of the cyclic and highest weight
representations of quantum algebras in physics, particularly in the theory of
quantum spin chains and the conformal field theories, please see :
V. Pasquier and H. Saleur, \np {330} {90} {523};
Z.-Q. Ma, {\sl Yang-Baxter Equation and Quantum Enveloping Algebras}, World
Scientific, Singapore, 1993.

\vfil

\bye